\documentclass[referee]{aa}  
\usepackage{graphicx}
\usepackage{txfonts}
\usepackage{epsfig}
\usepackage{natbib}
\bibpunct{(}{)}{;}{a}{}{,}
\begin{document}
   \title{174P/Echeclus: a strange case of outburst\thanks{Based on 
observations collected at the European Southern Observatory, Chile: ESO
N$^\circ$ 71.C-0092(A), 276.C-5042(B) and 278.C-5046(A).}}


   \author{P. Rousselot
          }

   \offprints{P. Rousselot}

   \institute{
         University of Franche-Comt\'e\\
         Observatoire de Besan\c con\\
         Institut UTINAM -- UMR CNRS 6213\\ 
	 BP 1615\\
	 25010 Besan\c con Cedex\\
	 France\\
	 \email{rousselot@obs-besancon.fr}
             }

   \date{}

 
  \abstract
   {More than ten Centaurs are now known to have cometary activity at
large heliocentric distance (i.e. $\simeq$ 5-13~AU). Among these objects,
174P/Echeclus which showed cometary activity at 13~AU from the Sun,
is a unique case, because of: (i) the amplitude of its outburst,
(ii) the source of cometary activity that appears distinct from 
Echeclus itself.}
   {This paper aims at investigating the physical conditions that have
led to this unusual outburst. The purpose is also to quantify this
phenomenon and to provide observational constraints for its modeling.}
   {We use observations from different telescopes, performed before,
during, and after the outburst. We performed the main observations 
on March 23 and 30, 2006, with the 8.2-m ESO Very Large Telescope and FORS~1 
instrument. They consist of visible images and spectra.}
   {Our main results are: (i) a cometary source distinct from Echeclus
itself that presents a brightness distribution compatible with a
diffuse source; (ii) a total dust production rate 
Q$_{dust}\simeq 86~kg.s^{-1}$ and a parameter $Af\rho\simeq 10,000$~cm; 
(iii) no emission lines (CN and C$_2$) can be detected
in the visible range; (iv) the upper limits for the CN and C$_2$ 
production rates are $\simeq 3.8\times 10^{25}$ and 
$\simeq10^{26}$ \ molecules.s$^{-1}$ respectively; (v) we detected 
no Echeclus' satellite before the outburst up to $M_R\simeq 26$; (vi) the upper 
limit for the object generating the coma is about 8~km in diameter; (vi) 
and we detected no cometary activity one year later, in March 2007.}
   {}


   \maketitle
%

\section{Introduction}
Centaurs, defined to
have their perihelion outside of Jupiter (5.2~AU) and semimajor axis 
inside of Neptune (30.0~AU), have unstable orbits and have been
extracted from the transneptunian objects (TNOs) population through 
perturbations by Neptune.
They represent probably an intermediate stage in the process that 
transforms a TNO to a short-period comet \citep{horner:2004}. 
Observations of cometary activity on these objects, always located at 
large heliocentric distances, are very useful for a better 
understanding of the physical relationship between Kuiper belt 
objects (KBOs) and comets. So far only a few Centaurs or TNOs 
classified as scattered disk objects (SDOs) have presented 
unambiguous evidence of cometary activity.

The first Centaur known to present cometary activity was Chiron. 
\cite{kowal:1977} discovered this object
and initially classified it as an asteroid. \cite{tholen:1988} suggested, on 
the basis of non-asteroidal brightness variations, that some kind 
of cometary activity could occur on this object. \citet{meech:1989} 
were the first authors to present a direct detection of Chiron's coma.
This object has a semimajor axis a=13.67~AU and a perihelion distance 
q=8.45~AU. It is considered a transition object between TNOs and 
Jupiter family comets. 

Some other objects with orbital elements that classified them as
Centaurs have been reported to have a cometary activity. They are
classified as comets. They are: 39P/Oterma (discovered in 1943); 
29P/Schwassmann-Wachmann~1 (discovered in 1927); C/2000~B4 (165P/LINEAR) 
\citep{kusnirak:2000}; C/2001 M10 (NEAT) \citep{lawrence:2001}; 
C/2001~T4 (166P/NEAT) \citep{pravdo:2001};
C/2004~PY$_{42}$ (167P/CINEOS), initially repertoried as 2004~PY$_{42}$ 
\citep{romanishin:2005}; P/2004~A1 (LONEOS) \citep{skiff:2004};
P/2005~S2 (Skiff); P/2005~T3 (Read) \citep{read:2005}; and 
174P/Echeclus (initially repertoried as (60558) 2000 EC$_{98}$, see below).

Two TNOs have also been reported by some authors to have cometary
activity. \cite{choi:2003} reported cometary activity for 
the SDO (29981) 1999~TD$_{10}$ but other observations with larger telescopes 
\citep{rousselot:2003, mueller:2004} detected nothing for
this object. Hainaut et al. (2000) presented a change in the lightcurve
of the TNO (19308) 1996~TO$_{66}$ that could possibly be interpreted as the
result of cometary activity.

This paper presents new observational data of 174P/Echeclus. This 
Centaur was discovered by the Spacewatch program on March 3, 2000
\citep{marsden:2000} and initially labelled (60558) 2000~EC$_{98}$. 
Table \ref{t:orbit} presents the orbital elements of this Centaur. 
Different observers have published obervational data -- mainly photometric in
nature -- related to this target. They have even searched for 
cometary activity without any success despite a very sensitive search of
up to magnitude 27/arcsec$^2$ \citep{rousselot:2005a,lorin:2007}.

\begin{table}[ht]
\caption{Orbital characteristics of Echeclus (from IAU Minor Planet Center).}
{ 
\label{t:orbit}
\begin{tabular}[t]{lllll}\hline   

a (AU)&e&q (AU)&Q (AU)&i\\
\hline
10.772&0.456&5.85&15.69&4.3\textsuperscript{$\circ$}\\
\hline
\end{tabular}
}
\end{table}

On December 30, 2005, a surprising cometary outburst was discovered with 
the 5-m Mount Palomar Observatory telescope \citep{choi:2006}.
This outburst corresponded to a change in the overall visual magnitude from
about 21 to about 14. At that time (60558) 2000~EC$_{98}$ was located at
13.07~AU to the Sun and was subsequently renamed with a cometary 
designation: 174P/Echeclus. Subsequent observations revealed
that the coma did not appear to be directly associated with the nucleus
\citep{choi:2006b}.

To study this outburst, we applied for Director's Discretionary
Time at the 8.2-m Very Large Telescope (VLT) of the European Southern
Observatory (ESO) in March 2006. 
The aim of these observations was to investigate in more detail 
this phenomenon by using optical imaging and spectroscopy. One year
later (March 2007) we conducted complementary observations with the
3.5-m New Technology Telescope of ESO.

In Sect. 2 the observational data are described.
Sects. 3 and 4 present our analysis of the images obtained, 
Sect. 5 the analysis of the spectroscopic data, and Sect. 6 is a discussion 
related to this phenomenon.
   
\section{Observations}

We first performed observations of 174P/Echeclus with the 8.2-m 
VLT of ESO with the Focal Reducer / Low Dispersion Spectrograph~1 
(FORS~1) instrument. 
This instrument is equipped with a  2$\times$2k Tektronix CCD,
thinned and anti-reflection coated (24~$\mu$m pixel). 
It was used both in direct imaging and standard resolution mode
and in longslit spectroscopy mode. These modes provide a scale of 
0.2~arcsec/pix.

A total of 3~hours of service mode observations were allocated to
our program. We split these observations to two different observing 
runs, in order to monitor the evolution of the coma. Table \ref{t:obs} 
presents the observing circumstances. The seeing 
was about 0.4~arcsec for the first observing night and 0.6 for the second one.
We obtained the long slit spectroscopy data with a 1.3~arcsec slit centered
on Echeclus and oriented in the motion direction (very close to the solar
direction). The spectral range covered
was 345-590~nm (chosen to cover the CN and C$_2$ Swan bands wavelengths)
with a spectral resolution of 600.

One year later we managed to obtain complementary observing time with
the 3.5-m New Technology Telescope (NTT) of ESO. These observations,
conducted in service mode with the direct imaging camera SUperb-Seeing 
Imager (SUSI~2), consisted in imaging Echeclus 
in the R, B, and V-band for one hour. Table \ref{t:obs} 
also presents the observing circumstances for these observations, which
we obtained, unfortunately, during a nonphotometric night.

\begin{table*}[ht]
\caption{Observing circumstances 
(R: heliocentric distance (AU); 
$\Delta$: geocentric distance (AU); $\alpha$: phase angle; mode: IMG (direct 
imaging) or LSS (long slit spectroscopy) ).}
\label{t:obs}
{ 
\begin{tabular}[t]{lllllllll}\hline   

UT start&R&$\Delta$&$\alpha$&Mode&Filter&Exposure time (s)&Instrument\\
\hline

2006 Mar. 23, 06:59:52&12.92&11.96&1.25\textsuperscript{$\circ$}&IMG&R&3$\times$120 s&FORS 1\\
2006 Mar. 23, 07:08:33&12.92&11.96&1.25\textsuperscript{$\circ$}&IMG&V&2$\times$145 s&FORS 1\\
2006 Mar. 23, 07:15:18&12.92&11.96&1.25\textsuperscript{$\circ$}&IMG&B&3$\times$145 s&FORS 1\\
2006 Mar. 23, 07:33:43&12.92&11.96&1.25\textsuperscript{$\circ$}&LSS&&4$\times$585 s&FORS 1\\
2006 Mar. 23, 08:23:51&12.92&11.96&1.25\textsuperscript{$\circ$}&LSS&&3$\times$585 s&FORS 1\\
2006 Mar. 30, 06:13:30&12.91&11.92&0.71\textsuperscript{$\circ$}&IMG&R&3$\times$120 s&FORS 1\\
2006 Mar. 30, 06:22:12&12.91&11.92&0.71\textsuperscript{$\circ$}&IMG&V&2$\times$145 s&FORS 1\\
2006 Mar. 30, 06:28:58&12.91&11.92&0.71\textsuperscript{$\circ$}&IMG&B&2$\times$145 s&FORS 1\\
2007 Mar. 24, 08:25:14&12.23&11.30&1.84\textsuperscript{$\circ$}&IMG&R&6$\times$125 s&SUSI 2\\
2007 Mar. 24, 08:40:59&12.23&11.30&1.84\textsuperscript{$\circ$}&IMG&V&6$\times$125 s&SUSI 2\\
2007 Mar. 24, 08:56:39&12.23&11.30&1.84\textsuperscript{$\circ$}&IMG&B&6$\times$125 s&SUSI 2\\
2007 Mar. 24, 09:12:30&12.23&11.30&1.84\textsuperscript{$\circ$}&IMG&R&4$\times$125 s&SUSI 2\\

\hline
\end{tabular}
}
\end{table*}

SUSI~2 is equiped with two 2k$\times$4k~CCDs providing a 
5.5'$\times$5.5' field of view. Because of the very small plate scale of the 
instrument (0.0805~arcsec/pixel) we used the 2$\times$2 binned mode.
To avoid any trailing due to the proper motion of the object the
exposure time was limited to 125~s, corresponding to a displacement of 
0.3~arcsec.
We obtained most of the images with a Bessel R filter, for which the
signal-to-noise is best.

\begin{figure}
\centerline{\psfig{file=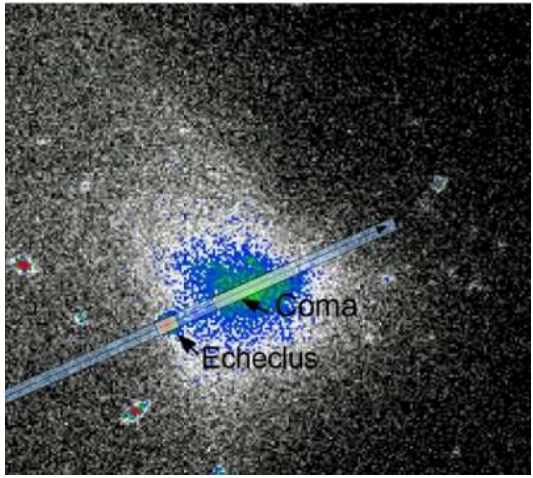,width=7.0truecm,angle=0,clip=}}
\caption{R-band image with the geometry of the long slit 
spectra obtained on Echeclus and its coma. The arrow indicates the direction
of the y-axis on the spectra.}
\label{f:spec}
\end{figure}

First, we preprocessed all the images obtained with FORS~1 (subtraction 
of the bias and division by a master sky flat-field). We specifically
processed the images corresponding to the spectra to extract 1D spectra. 
First, we calibrated in wavelength, thanks to wavelength calibration lamp images.
Second, we extracted two different 1D spectra from each image. The first one
corresponded to Echeclus itself (11 lines, corresponding to 2.2~arcsec
centered on Echeclus) and the second one to the center of the coma
(41 lines corresponding to 8.2~arcsec on the coma). We extracted the sky 
background from the same image (far from the nucleus, in the solar
direction) and we subtracted it from both spectra. Its intensity 
was dominant compared to the one of Echeclus and the coma. 
Figure\ref{f:spec} presents the geometry of the slit on the target and the 
corresponding area of the extracted 1D spectra.
We corrected the 1D spectra obtained above for atmospheric absorption
and calibrated in absolute flux thanks
to a spectrum of the spectrophotometric standard star 
EG274 observed twice during the observations. Finally, we median averaged
the 7 different spectra.

\begin{figure}
\centerline{\psfig{file=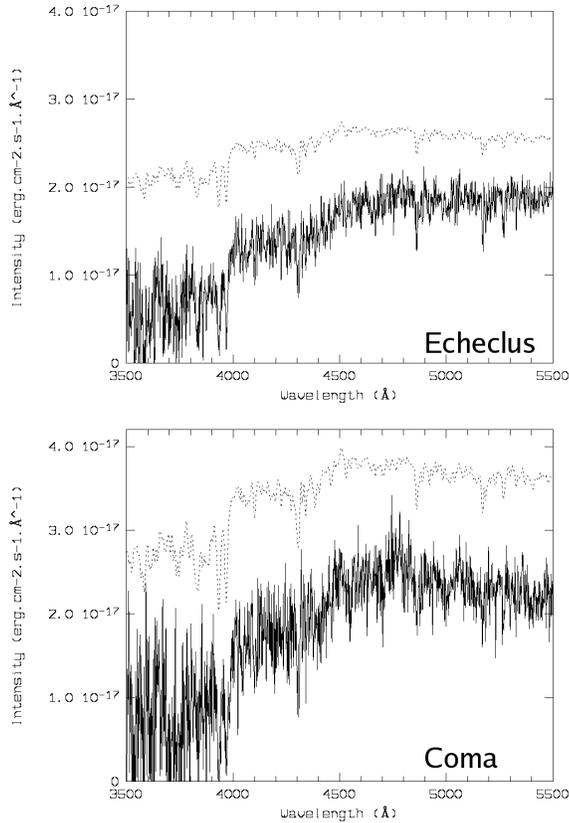,width=8.0truecm,angle=0,clip=}}
\caption{Resulting spectra obtained both for Echeclus (above) and 
the coma (below). In order to point out the influence of the scattered
solar spectrum a solar spectrum convolved with a similar instrument
response function has been added in both cases with dotted
lines (but has been shifted along the vertical axis for clarity).}
\label{f:spec2}
\end{figure}

The final step in the data processing consisted of subtracting a solar
spectrum convolved with a similar instrument response function
(FWHM=7\AA). The validity of the resulting solar spectrum
was checked by using the fraunhofer lines appearing in the coma
spectrum. We also checked the instrument response itself
with the wavelength calibration, lamp emission lines.
Because of the unknown reddening due to the light scattering by the 
dust particles in the coma this final step can only be correctly done 
in a limited part of the spectrum. For both spectral regions
corresponding to CN and C$_2$ emission lines (i.e. 3700-4000 and
4800-5200\AA) the solar spectrum was normalized with the observational
one before subtraction.
Figure \ref{f:spec2} presents an overall view of the averaged spectra
obtained both for Echeclus itself and the coma. 

We preprocessed the SUSI~2 images in a similar manner as the 
one used for FORS~1 images. We first subtracted a bias image computed as 
the average of five different bias images. A normalized flat-field,
corresponding to a median-averaged series of dome flat-fields, has been
used to perform the flat-field corrections.

\section{March 2006 Imaging}

\begin{figure*}
\centerline{\psfig{file=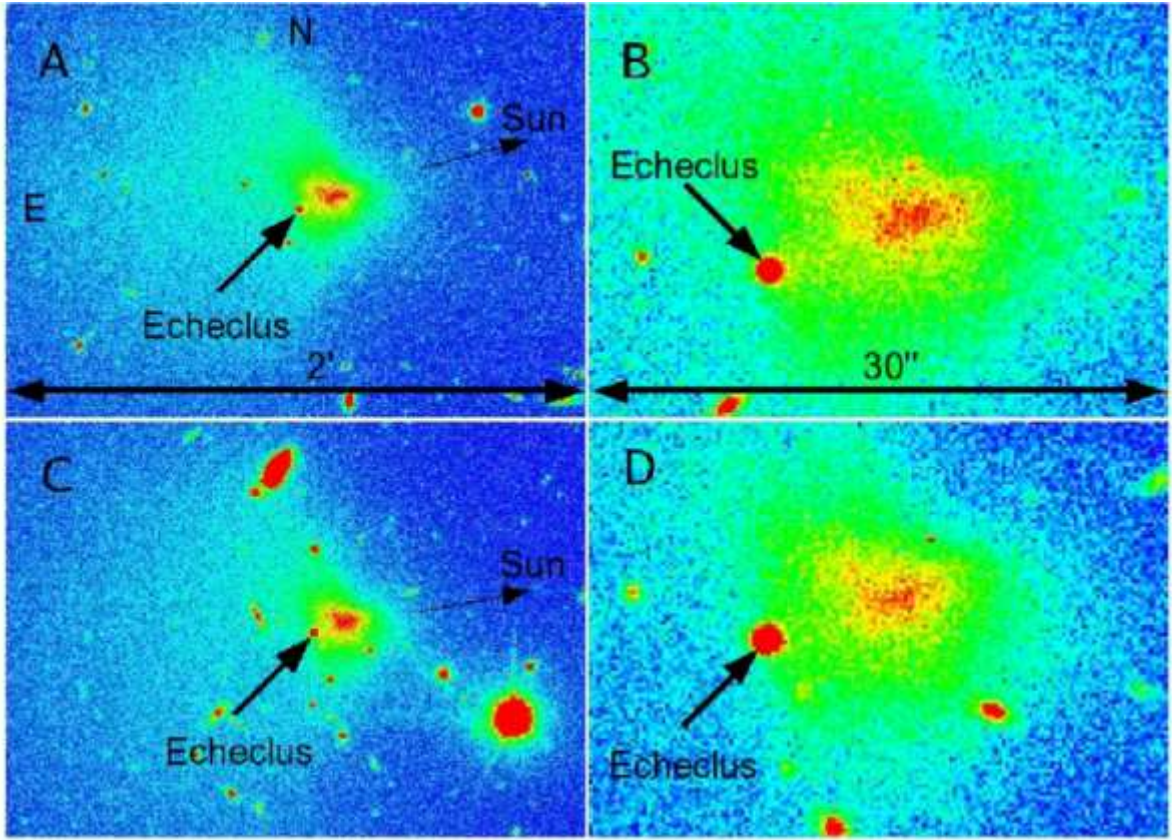,width=16.0truecm,angle=0,clip=}}
\caption{Echeclus images obtained on March 23 (top) and March 30 (bottom). 
Images B and D are a close-up view of images A and C, with a distinct color 
scale, to enhance the details near the center of the coma. Note that
Echeclus itself is distinct from the center of the coma.}
\label{f:imagesR}
\end{figure*}

The images obtained by FORS~1 clearly reveal that the coma is distinct
from Echeclus itself. Fig. \ref{f:imagesR} represents R-band images obtained
on March 23 and 30, 2006. It can be seen that the apparent distance between
the coma and Echeclus itself is roughly the same, i.e. about 60,000-70,000~km
($\simeq$8~arcsec).

\begin{figure}
\centerline{\psfig{file=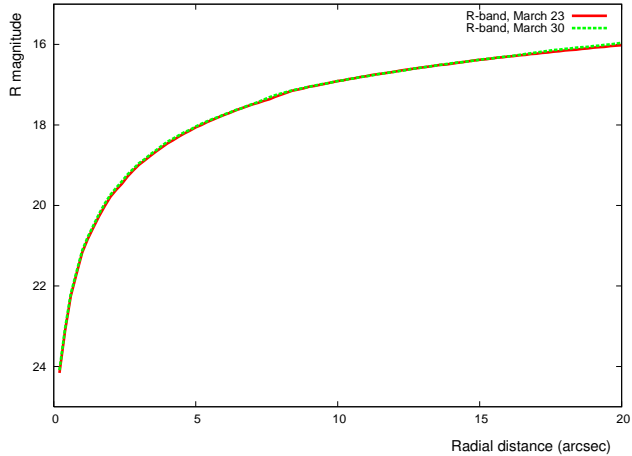,width=7.0truecm,angle=-90,clip=}}
\caption{Total magnitude of the coma (R-band) in function of the radial 
distance from the center of the coma. Comparison for the data obtained
on March 23 and March 30.}
\label{f:rmag}
\end{figure}

We computed the overall magnitude of the coma, by using standard 
star images and integrating all the flux in a square covering all the 
coma. This method provided acceptable results for the images obtained on 
March 23, because of the lack of bright stars superimposed on the coma. 
Table \ref{t:omag} presents our results. For the data 
obtained on March 30, the numerous stars and galaxies superimposed on
the image of the coma prevented us from using a similar method. We
compared the global apparent magnitude near the center of the coma 
for both observing nights (Fig. \ref{f:rmag}). Because there were no 
apparent differences, we conclude that there were no changes 
in the overall magnitudes on March 23 and March 30.

We have also computed the magnitude of Echeclus itself. This measure has been
conducted, despite the presence of the coma around the object, by using a small
radius to measure the flux and by measuring the sky background close to the 
object.The radius of the aperture used to measure the flux was 5 pixels 
and the sky background was measured in the 6 to 8 pixels annulus (with similar
values for the standard stars).
The results are presented in Table \ref{t:omag}. These values 
can be compared to previous ones obtained before the outburst. 
According to 
\citet{rousselot:2005a}, the average absolute magnitudes for the phase 
angle of the observations (1.25 and 
0.71$\textsuperscript{o}$) in the R-band are 9.30 and 9.25, respectively. 
Converted to apparent magnitude by taking into account the 
geocentric and heliocentric 
distances and by using the color indices, the average apparent magnitudes for 
March 23 and 30 should be, respectively: 20.24 (R), 20.75 (V), 21.51 (B),
and 20.19 (R), 20.70 (V), 21.46 (B). With a lightcurve amplitude of 
0.24$\pm0.06$~magnitude, these predicted apparent magnitudes are compatible with 
those appearing in Table \ref{t:omag}. We conclude that the nuclear
magnitude of Echeclus was unchanged during the 2005-2006 outburst: 
most or even all of the activity was associated with a separate source 
located within the coma condensation. 

\begin{table*}[ht]
\caption{Overall magnitude of the coma computed for the data of March 23.}
\label{t:omag}
{ 
\begin{tabular}[t]{llll}\hline   

band&coma magnitude&Echeclus magnitude (March 23)&Echeclus magnitude 
(March 30)\\

R&14.4$\pm$0.2&20.13$\pm$0.1&20.23$\pm$0.1\\
V&14.9$\pm$0.2&20.69$\pm$0.1&20.78$\pm$0.1\\
B&15.9$\pm$0.3&21.54$\pm$0.1&21.65$\pm$0.1\\
\hline
\end{tabular}
}
\end{table*}

We have also analyzed the images in term of the $Af\rho$ parameter. 
\citet{ahearn:1984} introduced this parameter, which permit us to 
quantify the cometary activity level in a manner not too sensitive 
to the parameters used to quantify dust production rate (e.g. 
albedo or dust grain density). 
In this expression, $A$ represents the bond albedo, i.e. the ratio 
of the total scattered radiation to the total incident radiation 
(for isotropic scatterers $A=4p_v$ if $p_v$ is the geometric albedo);
$f$ is the filling factor of grains within the field of view and 
$\rho$ is the projected cometocentric distance. The filling factor
is the total cross section of grains within the field of view 
divided by the area of the field of view.
The $Af\rho$ parameter supposes 
a steady state coma created by a nucleus with a $1/\rho$ variation of 
the dust quantity (or number of dust particles) 
integrated along the line of sight 
and is supposed to be, more or less, independent of the 
aperture size used to measure it. We have used the formula presented in 
\citet{lorin:2007} for computing the $Af\rho$ parameter (and the standard 
star images) in the three different bands\footnote{We have used 
some slightly different magnitudes
for the Sun than the one mentionned in this paper 
(see www.ucolick.org/~cnaw/sun.html),
measured through R, V and B Bessel filters: -27.12 (R), -26.78 (V), 
and -26.14 (B). 
These magnitudes lead to the q values \citep{lorin:2007}
equal to $1.53\times10^{15}$, $1.32\times10^{15}$, and $1.31\times10^{15}$ 
respectively}. The total $Af\rho$ parameter (when integrating for the 
whole coma, with the R magnitude presented in Table \ref{t:omag}) can be 
estimated to about 10,000~cm. This value is similar to the one
computed by \citet{bauer:2006} on the basis of observations performed
in R-band at the end of February 2006 during coordinated optical-infrared
(Spitzer) observations of Echeclus. 

\begin{figure}
\centerline{\psfig{file=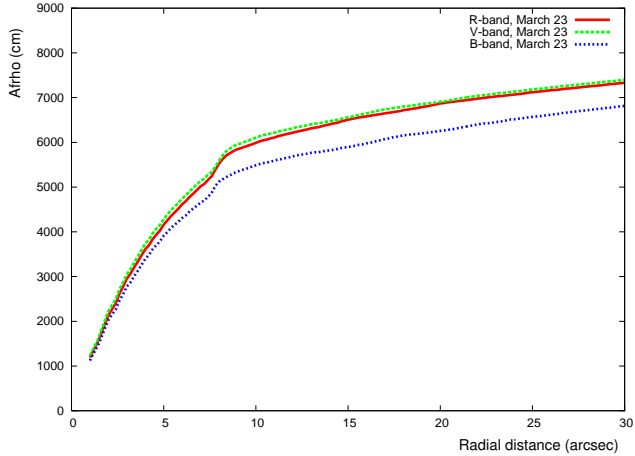,width=7.0truecm,angle=-90,clip=}}
\caption{$Af\rho$ parameter vs radial distance for the three different bands
and the images obtained on March 23, 2006. Each profile is the average of
the profiles obtained for the different images.}
\label{f:afr1}
\end{figure}

Figure \ref{f:afr1} presents the resulting profiles. 
The examination of these profiles reveals that the $Af\rho$ parameter is 
not constant with the radial distance but increases with it. 
The increase of $Af\rho$ with
the radial distance could be interpreted -- if observed only for March 23 -- 
by a decrease of the dust production rate with time, the regions observed 
close to the center of the coma containing ``younger'' dust particles. The 
similar profiles obtained one week later (see Fig.~\ref{f:rmag}) 
permit to discard this 
hypothesis and lead to the conclusion that this variation of the $Af\rho$ 
parameter with the cometocentric distance is due to a variation in
the intensity that does not follow the $1/\rho$ intensity variation 
assumed by a
nucleus releasing dust particles. The more plausible explanation for such a 
behavior is a steady state process (at least at the timescale of our 
observations) created by a \textit{diffuse} source of dust. Such an 
interpretation is coherent with the general view provided by the different 
images.

In the case of a $1/\rho$ intensity variation for the intensity profile 
$I(\rho)$ the $Af\rho$ parameter
is more or less independent of $\rho$ because $Af\rho=k\times F_{com}/\rho$ and 
$F_{com}=\int_0^\rho k'2\pi rI(r)dr=\int_0^\rho k'2\pi rk''/r dr=2\pi k'k''\rho$
(where $k$, $k'$ and $k''$ are some constants and $F_{com}$ is the coma flux). 
In the case of our observations
$Af\rho$ increases with $\rho$ because $I(\rho)$ varies as $\rho ^{-x}$ with
$x<1$ leading to $F_{com}=cte\times\rho^y$ with $y>1$.

It is possible to use the global $Af\rho$ parameter to derive an approximate
dust production rate by using this equation with  $Af\rho=100~m$ 
(see \citet{lorin:2007} for more details):

\begin{equation}
Q={2 Af\rho  a_{gr}v_{gr}d\over 3p_v}
\label{eq:Q}
\end{equation}

We used the following parameters: grain radius $a_{gr}=0.5\times 10^{-6}$\ m,
volumetric mass $d=1000$~\ kg.m$^{-3}$ and geometric albedo 
$p_v=0.05$. For the dust ejection velocity $v_{gr}$ we used the
formula $v_{gr}=465R^{-0.5}$\ m.s$^{-1}$, with $R$ the heliocentric distance
expressed in AU \citep{delsemme:1982}.
These parameters lead to a total dust production rate of 
Q$\simeq$86~kg.s$^{-1}$. Such a production rate 
can be compared to the upper limit obtained with our April 2001 observations 
\citep{lorin:2007}, which was 28~cm for Af$\rho$ 
and 0.45~kg.s$^{-1}$ for the dust 
production rate Q (with similar parameters to the one mentioned above). The
dust production rate has increased by at least a factor of 
$\simeq$200 between 2001 and 2005 outburst. The fact that
the dust is created by a diffuse source impacts the interpretation
of the $Af\rho$ parameter in terms of the dust production rate. The above
formula is based on a $1/\rho$ distribution. With a flatten
dust distribution the dust production would be slightly smaller
for a same $Af\rho$ value. The $Q$ value provided above should be 
regarded, consequently, as an upper limit.

\begin{figure*}
\centerline{\psfig{file=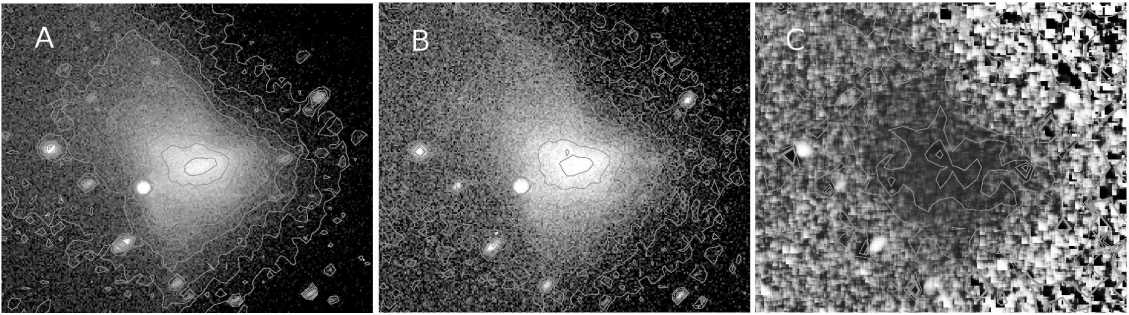,width=15truecm,angle=0,clip=}}
\caption{R-band and V-band images, and their ratio. 
A: average image of the three R-band images obtained on March 23. 
B: average image of the two
V-band images obtained on March 23. C: ratio of the R-band to the
V-band image. The scale
is similar for the three images (50$\times$41 arcsec) and the 
images A and B are corrected for the sky background, the exposure times, and 
the difference in the relative sensitivity. The image C has been smoothed with
a gaussian filter of 7$\times$7 pixels size. Black represents dust 
that is relatively blue, while white represents relatively redder dust. 
The ratios appearing in image C are in the range $\simeq$1.3 (center of 
the coma) to $\simeq$1.5 outside the coma.}
\label{f:color}
\end{figure*}

We have also tried to compute color ratios. We have subtracted 
the sky background for R-band and V-band data, divided by the exposure
time and corrected for the zero point before computing the ratio.
Figure \ref{f:color} presents
the ratio of the R-band to the V-band data (the one with the best 
signal-to-noise). It can be seen that the redenning increases with 
cometocentric distance. This color change is probably indicative of
a grain fragmentation process. According to the Mie
scattering theory, the light scattered by
dust particles depends on the grain radius and the wavelength. If the
grain size changes with cometocentric distance, the light scattered
by the dust will be affected in a different manner at different wavelengths.
 
\section{March 2007 Imaging}

\begin{figure}
\centerline{\psfig{file=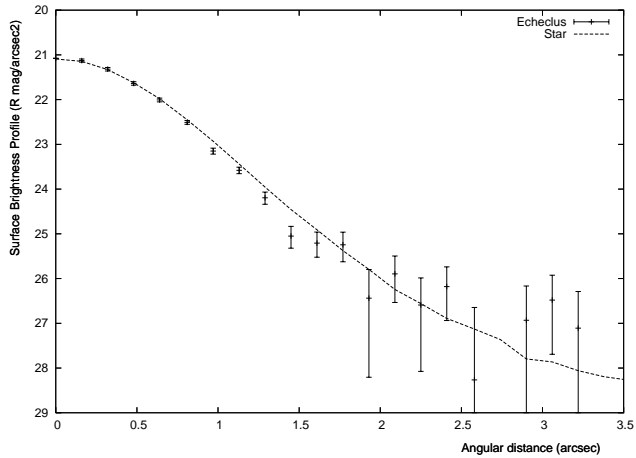,width=7.0truecm,angle=-90,clip=}}
\caption{Surface brightness profile of Echeclus compared to the one 
of a star. These profiles are derived from the 2007 observations
conducted with SUSI~2 and correspond to a quiescent state}.
\label{f:sbp}
\end{figure}

The images obtained with SUSI~2 on March 24, 2007 did not allow us to detect
any cometary activity. Because no photometric standard star could be observed
during the same night (nonphotometric night) we have used the predicted
Echeclus average magnitude for calibrating the profile with absolute
magnitudes. We base this prediction on the phase curve published by
\cite{rousselot:2005a}, and the heliocentric and geocentric distances, as
well as the phase angle during the observations.
Figure \ref{f:sbp} presents the surface brightness
profile of Echeclus obtained when all the R-band images are co-added
(after centering). We compare this profile to the one of a star
apprearing in the same field of view and adjusted in intensity. No
differences can be detected up to the sky background, i.e. up to
R$\simeq$27/arcsec$^2$. We have used the method mentioned by \cite{jewitt:1984}
to derive an order of magnitude of the maximum $Af\rho$ parameter that
can be derived from these data. We have used the $R_{max}$
magnitude where the surface 
brightness profile reaches the sky background, i.e. about 27/arcsec$^2$. 
From this magnitude we have derived a lower limit for the coma magnitude,
if it exists, by using the formula: 
$R_{coma}\simeq-2.5Log_{10}(2\pi r^2)+R_{max}$. In this formula,
$r$ represents the diameter corresponding to the $R_{max}$ magnitude,
i.e. 5~arcsec.
These values lead to an upper limit for $Af\rho$ equal to about 75~cm. 
This upper limit can be compared to the one measured one year before
(10,000~cm for the R-band, see above). The ratio is about 130.
By using similar parameters and formulae as the one mentionned
in Sect. 3 we derive an upper dust production rate 
$Q_{max}\simeq 0.6$~kg.s$^{-1}$.

\section{Spectra}

In the range covered by our spectra different
emission bands corresponding to different radicals could be 
observed.
Among these emission bands the more intense are CN (3880~\AA) and C$_2$ (Swan
bands with the strongest one having a bandhead at 5165~\AA). Because of
the large heliocentric distance CN is expected to be dominant,
if some emission bands are apparent. This is due to the fact that the ratio
CN/C$_2$ increases with heliocentric distance \citep{ahearn:1981} 
because CN is
less sensitive to the heliocentric distance. So far CN is also the only
emission band detected in the visible range at large heliocentric distance
(on Chiron, see \cite{bus:1991}).

\begin{figure}
\centerline{\psfig{file=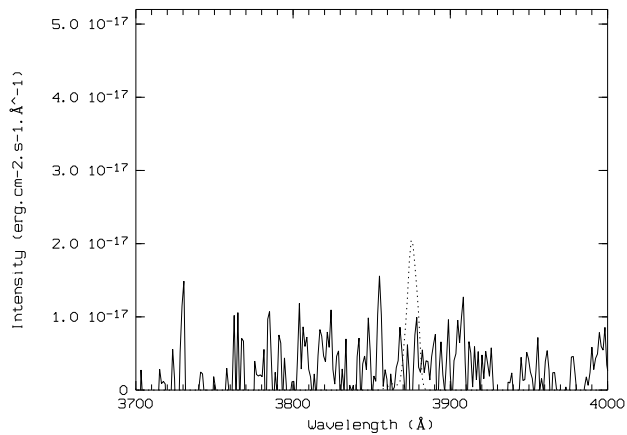,width=6.0truecm,angle=-90,clip=}}
\caption{Observational spectrum of the coma obtained in the wavelength
region corresponding to CN emission band (the solar continuum being 
subtracted). A theoretical CN spectrum
convolved with a same response function has been added in dashed line.}
\label{f:cn}
\end{figure}

\begin{figure}
\centerline{\psfig{file=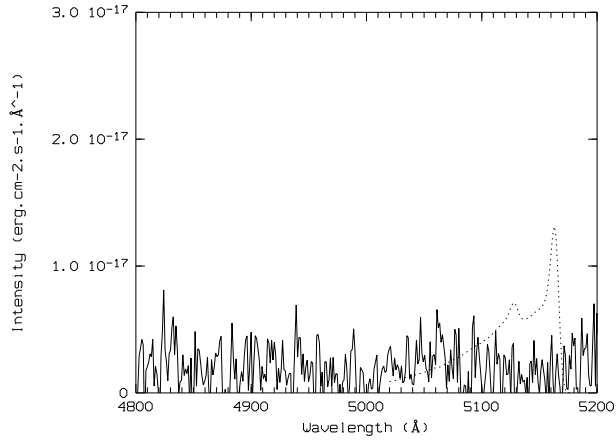,width=6.0truecm,angle=-90,clip=}}
\caption{Observational spectrum of the coma obtained in the wavelength
region corresponding to C$_2$ emission bands (the solar continuum being 
subtracted). A theoretical C$_2$ spectrum
convolved with a same response function has been added in dashed line.}
\label{f:c2}
\end{figure}

We have searched for C$_2$ and CN emission bands. Figure \ref{f:cn} 
presents the observed spectra obtained in the 3700-4000~\AA\ range
(CN band), and Fig. \ref{f:c2} presents the 4800-5200~\AA\ region 
(C$_2$ bands) after subtraction of the 
solar spectrum. A theoretical spectrum of both CN and C$_2$ emission bands
has been superimposed on these spectra.

We computed the CN spectrum by using the model described in 
\cite{zucconi:1985}. We computed the C$_2$ spectrum with the model 
described in \cite{rousselot:2000} with transition moments 
$|D_{a-X}|^2=|D_{c-X}|^2=3.5\times 10^{-6}$\ atomic unit. 
Both spectra were computed for similar heliocentric 
distance and velocity (-3.097~km.s$^{-1}$) and
convolved with an instrument response function similar to that
of FORS~1 in the mode used during our observations (7\ \AA\ of FWHM).
As we've seen, no CN nor C$_2$ emission lines are apparent. It is
only possible to derive an upper limit for both of these species.

In the case of CN the brightest possible CN emission band can be 
estimated to have an intensity equal to about 
$2.0\times 10^{-17}$\ erg.cm$^{-2}$.s$^{-1}.$\AA$^{-1}$ (corresponding 
to a 4-sigma detection level). Taking 
into account the bandwidth this peak corresponds to 
1.6$\times10^{-16}$\ erg.cm$^{-2}$.s$^{-1}$ of the maximum possible flux 
for the CN band. Taking into account the fluorescence efficiency factor 
g of the CN band computed for such heliocentric distance and velocity 
(4.4$\times 10^{-4}$~photon.molecule$^{-1}$.s$^{-1}$), and 
the geocentric distance
(11.962~AU), this leads to a maximum number of CN radicals inside the
slit of about 2.9$\times 10^{28}$.

This number can be related to a production rate. Such a calculation
can only be approximate, because of the unknown parameters in such a 
model. The production rate would probably not be constant with time
because it is an outburst, the velocity of the daughter radical is 
unknown, and the phenomenon is probably not isotropic. We have used a
Haser's model, nevertheless, to make a rough estimate of
the production rate. For this model, which supposes a steady
state process, we have used the following parameters: (i) a 
parent molecule scalelength $l_p=2\times 10^4$~km (at 1~AU); (ii) a 
daughter scalelength $l_p=3\times 10^5$~km (based on \cite{waniak:2007}); 
and (iii) a CN average expansion
velocity equal to the dust ejection velocity, taken equal to 
465.R$_h^{-0.5}$ \citep{delsemme:1982}, R$_h$ being the heliocentric 
distance expressed in AU.
With these parameters and the slit size, we compute that the maximum
possible CN parent molecule prodution rate is
Q$_{CN}\lesssim 3.8\times 10^{25}$ molecules.s$^{-1}$.

This result can be compared to the other procution rates of CN parent 
molecules measured in other comets. For example it is similar to the
one observed in Halley's comet in November 1985 \citep{almeida:1992} or,
to consider a more similar case, to the detected CN on Chiron 
\citep{bus:1991}. In both cases the absolute overall magnitude of the object was
fainter than the magnitude of Echeclus. For Echeclus, the overall magnitudes 
presented in Table \ref{t:omag} lead to an absolute magnitude of
V$\simeq$4, brighter than Halley's comet's
absolute magnitude (close to 5 in November 1985) and the one
of Chiron (close to 6 when CN was detected). 
The CN radicals, if present in the coma of Echeclus, are consequently
underrepresented compared to the dust when compared to the ``usual''
comets. 

For the C$_2$ radical similar calculations lead to 
Q$\lesssim 10^{26}$~molecules.s$^{-1}$. This value is based on a 
g factor equal to 5.7$\times$10$^{-4}$\ photon.s$^{-1}$.molecule$^{-1}$
for the Swan bands sequence $\Delta v=0$ at the heliocentric distance 
of Echeclus (see Fig. \ref{f:c2}) and an overall maximum
flux of 6.5$\times 10^{-16}$\ erg.cm$^{-2}$.s$^{-1}$ (based on a 
maximum intensity of $1.3\times 10^{-17}$\ erg.cm$^{-2}$.s$^{-1}.$\AA$^{-1}$).

Like the CN radical, the value of Q$_{max}$ is similar to the values 
observed in Halley's comet from November 1985, ground-based observations
\citep{almeida:1992}, which had a fainter absolute magnitude than
for Echeclus. Consequently the gas-to-dust ratio is 
much lower than in the cometary comae (if molecular species 
are present in the Echeclus' coma).

\section{Discussion}

The event that happened to Echeclus looks like a cometary outburst.
Some other planetary bodies are now known to have suffered a 
similar event at large heliocentric distance (above 5~AU), nevertheless
this outburst presents unusual characteristics that deserve a more detailed
discussion: (i) it is unique by its amplitude (about 7 magnitudes); 
(ii) the heliocentric distance is 
important for this type of event (12.9~AU); (iii) the coma appears
distinct from the object itself for a long time (at least several months); and
(iv) the brightness distribution is compatible with a diffuse source.
Two different hypotheses might explain these characteristics:
(i) the coma is created by a fragment ejected from Echeclus, or (ii) it
is created by a previously unknown satellite.

We examine the first hypothesis (a fragment ejected from Echeclus)
in more detail. First, no point-like object appears in this coma. From this
observational fact, it is possible to derive an upper limit for a point-like
object that would be responsible for this cometary activity. We have added
the image of a point-like object in the center of the coma with a two-sigma
intensity (compared to the standard deviation of the intensity in this
area). This intensity corresponds to a magnitude of $\simeq$25 (R-band).
This  magnitude, with a geometric albedo of 0.04, corresponds to
a diameter of 8.3~km for the considered geocentric and heliocentric 
distances. With a 1-sigma detection level and a geometric albedo of 0.1 
the upper limit for the diameter would be 3.6~km. 

The diameter of Echeclus is estimated to be 83.6$\pm15$\ km, and its visual 
geometric albedo to 3.83$^{+1.89}_{-1.08}$\% \citep{stansberry:2007}. 
These upper limits for a fragment seems to be realistic and of the order of
magnitude for a cometary nucleus. The problems raised by a fragment are:
(i) the fact that the surface brightness variation does not seem compatible
with a point-like source (it does not follow a 1/$\rho$ law); and (ii) the 
event responsible for the ejection of this fragment. The point (i)
probably implies a fragmentation process, as already pointed out by the
color changes with cometocentric distance. Perhaps the matter ejected
from the nucleus is more similar to a swarm of dust particules
than to an ice fragment.

We also examine the hypothesis of a previously unknown satellite 
more carefully. With reasonable parameters, i.e. a diameter 
of 83.6~km, a volumetric mass of 1000~kg.m$^{-3}$, and
a semi-major axis of 60~000~km (the minimum possible distance, because
it is the projected distance on the sky), the third Kepler's law implies
an orbital period P$\simeq$20~years. This period could be shorter in
the case of a highly eccentric orbit with a smaller semi-major axis, but
such an eccentricity would be very unusual. A larger semi-major
axis cannot be excluded and would lead to a longer period. In any
case, the order of magnitude of such a period is compatible with an
apparently motionless object on a timescale of a week.

\begin{figure}
\centerline{\psfig{file=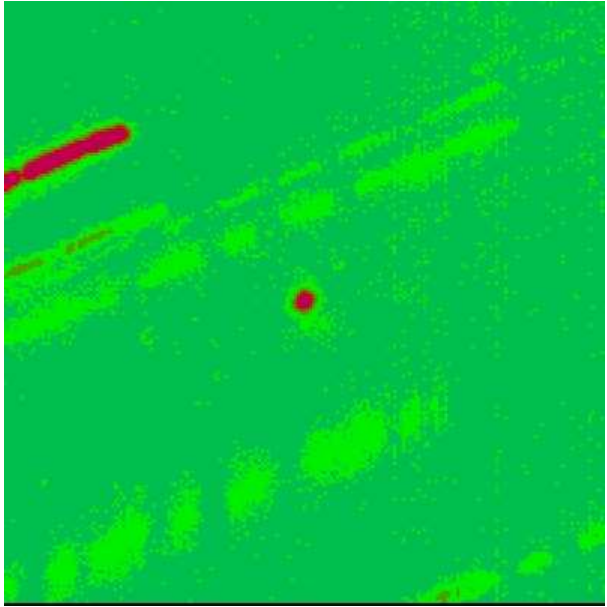,width=8.0truecm,angle=0,clip=}}
\caption{Archive data of Echeclus obtained with the ESO 3.6-m telescope
on April 10, 11 and 12, 2003. This image is a co-addition of all these
data which correspond to a total integration time of 7.5~hours. North
is up, East to the left, and the image scale is 1$\times$1~arcmin.}
\label{f:t360}
\end{figure}

It is also possible to examine the data obtained before the outburst.
The best are those obtained with the ESO \makebox{3.6-m} telescope on 
April 10, 11 and 12, 2003 (see \cite{rousselot:2005a} for more details).
We have coadded all these data (total integration time of 7.5~hours)
and could not find any evidence of a satellite up to $m_R\simeq 26$
(Fig. \ref{f:t360}). The apparent magnitude at the time of the
observations corresponds to an absolute magnitude of 14.5, i.e. to an 
upper diameter limit of 7~km (with a R geometric albedo of 0.04). 
Such an upper limit is compatible with the one derived from the coma itself.

If the cometary activity presently observed is created by a 
satellite, two issues remain unexplained: (i) what is the origin of the
outburst (collision ?); and (ii) why does it appear as a diffuse source ?
Other investigators \citep{weissman:2006} have also pointed out the apparent
random motion of this source on the timescale of several months, on
the basis of their own observations. If such a random motion is confirmed
it would exclude the satellite hypothesis.

Finally, the more realistic explanation remains that Echeclus has ejected
a fragment. The reason for this ejection remains
unclear, and is still to be investigated in more detail. This fragment
has probably suffered a disintegration process.

The mechanism responsible for the outburst is probably not a simple
impact that would have thrown off dust particules 
because it would not have lasted several months and some changes 
would have been apparent
at the timescale of a week. As for other similar events observed at large
heliocentric distances, a more complex process probably occured. It can 
be either a CO-driven activity or driven by an amorphous $\rightarrow$
crystalline phase transition for water ice. The onset of activity was
probably triggered by an unknown external phenomenon because it did 
not occur at the smallest heliocentric distance. It is also important 
to point out that the 2007 observations, which did not permit 
to detect any activity, were performed at a smaller heliocentric distance.

\section{Conclusions}

The Centaur (60558) Echeclus, renamed 174P/Echeclus after the discovery of
an important cometary outburst, has been observed with FORS~1 
at VLT. The main conclusions of our observations are: 

$\bullet$ The source of cometary activity appears distinct from
Echeclus itself (about 8~arscec, corresponding to a projected distance
of about 60,000-70,000~km), and stable at the timescale of a week.

$\bullet$ The brightness distribution of this source does not follow
that of a cometary coma created by a point-like source (cometary
nucleus). It look likes a diffuse source.

$\bullet$ No cometary emission lines can be detected in the visible
range (345-590~nm). The upper detection limit computed for the
CN and C$_2$ emission bands, when compared to the overall brightness
of the phenomenon, leads to a significantly lower gas-to-dust ratio
for Echeclus when compared to the comets observed close to the Sun.

$\bullet$ The global Af$\rho$ parameter is equal 
to about 10,000~cm (R-band, observations of March 23, 2006), 
corresponding to a global
magnitude of 14.4$\pm$0.2. These parameters do not significantly
change for the observations conducted on March 30, 2006.

$\bullet$ Our observations performed one year later with
SUSI~2 at NTT on March 24, 2007,
did not permit to detect any cometary activity, despite a smaller
heliocentric distance. These observations lead to 
$Af\rho\lesssim 75$ cm.
The phenomenon observed in 2006 seems to be, up to now, unique.

\section*{Acknowledgments}

The author is grateful to Jean-Marc Zucconi for modeling the CN spectrum.

\bibliographystyle{aa}
\bibliography{references_aa.bib}

\end{document}